\documentclass[iop]{aastex}

\newcommand{{\HII}}{H\,{\sc ii}}

\newcommand{\um}{\,$\mu$m}
\newcommand{\kms}{\,km\,s$^{-1}$}

\shorttitle{[NeII] in NGC 4194}
\shortauthors{Beck et al}

\begin{document}

\title{Ionized Gas Kinematics at High Resolution IV: Star Formation and a Rotating Core in the Medusa (NGC 4194) \\
   }

\author{Sara C. Beck\altaffilmark{1,2}, John Lacy\altaffilmark{2,3}, Jean Turner \altaffilmark{4}, Thomas Greathouse \altaffilmark{2,5}, Susan Neff \altaffilmark{6}}

\altaffiltext{1}{School of Physics and Astronomy, Tel Aviv University, Ramat Aviv ISRAEL 69978}
\altaffiltext{2}{Visiting Astronomer at the Infrared Telescope Facility, which is operated by the University of Hawaii under Cooperative Agreement no. NNX-08AE38A with the National Aeronautics and Space Administration, Science Mission Directorate, Planetary Astronomy Program.}
\altaffiltext{3}{Department of Astronomy, University of Texas at Austin, Austin TX 78712}
\altaffiltext{4}{Department of Physics and Astronomy, UCLA, Los Angeles, CA 90095-1547}
\altaffiltext{5}{Southwest Research Institute, San Antonio TX 78228-0510}
\altaffiltext{6}{ NASA-Goddard Space Flight Center, Greenbelt MD 20771}
\email{becksarac@gmail.com}

\begin{abstract}
NGC 4194 is a post-merger starburst known as The Medusa for its striking tidal features.  We present here a detailed study of the structure and kinematics of ionized gas in the central 0.65 kpc of the Medusa.  The data include radio continuum maps with resolution up to 
$0.18\arcsec$ (35 pc) and a $12.8\mu$m [NeII] data cube with spectral resolution $\sim4$\kms: the first {\it high resolution, extinction-free} 
observations of this remarkable object.  The ionized gas has the kinematic signature of  a core in solid-body rotation.  The starburst has formed a complex of bright compact \HII~regions, probably excited by deeply embedded super star clusters, but none of these sources is a convincing candidate for a galactic nucleus. The nuclei of the merger partners that created the Medusa have not yet been identified.    \end{abstract}

\keywords{galaxies: individual(NGC 4194)--galaxies: starburst--galaxies: star formation--galaxies: kinematics and dynamics}

\section{Introduction}

The most luminous starbursts appear to be caused by major mergers
of galaxies \citep[e.g.,]{1988ApJ...328L..35S} and are rare because major mergers are infrequent.  Minor mergers create 
LIRGS; slightly less luminous but much more common, and important to galaxy 
evolution due to their sheer numbers  \citep{KA10}.
Minor mergers, like major mergers, appear to instigate intense star formation in compact regions and create concentrated 
sources.  But minor mergers evolve differently from major mergers, so the 
characteristics of star formation are likely to be different.  How does star formation develop in a minor merger? 
 How will these starbursts evolve and how do they affect their surroundings as they
 contribute to the newly formed galaxy?  
   To answer those questions we need to understand the structure and kinematics of the gas ionized by the embedded stars with the highest possible spectral and spatial resolution. We consider
   here the case of NGC~4194, ``The Medusa", a galaxy with distorted morphology
   and luminous star
   formation that appears to be the result of an unequal 
merger between an elliptical and a smaller spiral.

This is the fourth paper
in a series on high resolution spectroscopy of intense extragalactic star formation sources in the middle infrared \citep{BT10,BT12, BT13}.  We use mid-infrared lines of metal ions to trace the kinematics and spatial distribution of ionized gas in Galactic {\HII} regions and in starburst galaxies \citep{{AC95},{ZH08},{BT13}} because they are little affected by extinction and  permit us to attain true spectral resolution, including thermal effects, much higher than is possible with any hydrogen line.  
The 'Medusa galaxy',  NGC 4194 (a.k.a. Arp 160 and Mkn 201) is a starburst galaxy at 39 Mpc %%
distance (1\arcsec = 190 pc) with striking tidal features; it has been variously classed as Magellenic, Sm(pec)  and BCG.   The central kpc of NGC 4194 is a powerful infrared source with 60\um/100\um $\approx 1$, typical of starburst heating.  Spitzer IRS spectra (from the Heritage Archive) show a spectrum typical of a starburst-dominated galaxy, with little if any contribution from an AGN.   NGC 4194 is believed to be a merger remnant, with the favoured history being a small gas-rich spiral falling into an elliptical four times it mass \citep{MA08}. The central region of NGC 4194  is the site of intense star formation: \citet{WE04} and \citet{HA06} find numerous bright optical and UV knots that they identify as very young globular cluster precursors.   But the structure of this central starburst has not been probed. The observations are either too low in spatial resolution (e.g. the 2Mass survey)  or too affected by the deep extinction \citep{HA04} to determine the structure.    \citet{BE05} mapped the center of NGC 4194 at 21 cm with sub-arcsecond resolution  and found two compact sources separated by only 0.35\arcsec~, which they identify with the galactic nucleus, and  which  \citet{MA08} appear to tentatively accept as the nuclei of the progenitor galaxies, but the nature of these source and their relation to the merger history have not been explored.

We report here on radio continuum maps of NGC 4194 with sub-arcsecond spatial resolution and on spectra of the 12.8\um~emission line of [NeII] with velocity resolution better than 5\kms. 
 The radio maps are at short cm wavelengths, which are sensitive to both the thermal emission of {\HII} regions and the non-thermal emission of SNR.     The fine-structure line of $Ne^+$ at 12.8\um~ is usually one of the strongest mid-infrared emission lines in {\HII}~ regions and is a preferred  kinematic probe because of its low susceptibility to thermal broadening; in a  $T_e =7500 K$  {\HII} region the FWHM from thermal effects will be $4.1$\kms~ compared to  $18$\kms~ for $H^+$.  
\section{Observations}
\subsection{TEXES [NeII] Data Cube}
NGC 4194 was observed at the NASA Infrared Telescope Facility on Mauna Kea, Hawaii, on the night of 2 February 2013, with the TEXES spectrometer \citep{LA02}.   TEXES on the IRTF has a seeing and diffraction-limited beam of $\sim1.4\arcsec$. The observations  were taken in the high-resolution mode, with a plate scale~$0.36\arcsec\times0.9~\rm km~s^{-1}$ per pixel and spectral resolution $\sim3$\kms .   Ceres was the telluric comparision source and the wavelength scale was set from atmospheric lines.  The slit was 29-pixel long and $1.32\arcsec$ wide; it was set North-South and scanned across the galaxy in R.A. steps of $0.7\arcsec$ to generate data cubes
with Nyquist spatial sampling.   The total observation time was 2.23 hours. 
\subsection{Radio Maps} 
NGC 4194 was observed at the NRAO Very Large Array (VLA)\footnotemark~ in program AN095.  A array observations at 6  and 3.6 cm (C and X bands)  were obtained on 17 November 2000 and  B array observations at X band on 19 March 2001.   The data were calibrated and reduced with AIPS.  Maps were produced with varied weightings and beam sizes; the observing and reduction parameters are given in Table 1.   
\subsection{Archival Data}
\subsubsection{Radio}
We obtained archival data in the radio and infrared regimes from the VLA and from Spitzer.  NGC 4194 was observed at 2 cm (U band) in the A array for program AE47 and at  6 cm with the B array  in program AS286.    Note that the resolution of these B array observations is on the order of  the diffraction-limited TEXES beam.     The parameters of these maps are given in Table 1. 
\footnotetext{The National Radio Astronomy Observatory is a facility of the National Science Foundation operated under cooperative agreement by Associated Universities, Inc.}
\subsubsection{Infrared Spectra}
NGC 4194 was observed by Spitzer as part of the IRS Standard Spectra program and the results are in the Spitzer Heritage Archive.  It is dominated by the low-excitation lines of [NeII] and [SIII] $18.7$\um.  All the 
lines appearing in the  $9.9-19.6$\um~SH and  $18.7-37.2$\um~ LH modules are in the ratios expected of a starburst galaxy \citep{ST02};  there is no spectral signature of an AGN. 
\section{\HII~Regions in the Central Kpc of NGC 4194}
\subsection{Radio Continuum}
Optical and UV observations of NGC 4194 have detected multiple star-forming knots near the nucleus \citep{{WE04},{HA06}}, but because of heavy dust obscuration these observations did not penetrate the nucleus itself.  Published infrared and radio observations are either of low resolution or at a wavelength not sensitive to the thermal emission of young starforming regions.  The radio observations we report here are the first high-resolution, obscuration-free images of star formation activity in the nucleus of NGC 4194.   We first discuss the spatial distribution and structure of the nucleus, and then the nature of the radio sources. 
\subsubsection{Spatial Distribution of Radio Emission}
Maps of NGC 4194 at 6 and 3.6 cm are shown in Figure 1, and a 2 cm map, converted from the B1950 to J2000 system, is in Figure 2.  This is the first complete and extinction-free 
high resolution view of 
ionized
gas structure near the nucleus in NGC 4194. The radio continuum in the central $kpc$ of NGC 4194 has a complex structure which our highest resolution maps barely resolve.  The central source\footnotemark~ at $\alpha=12^h14^m9.68^s,\delta=+54^{\circ}31'35.8''$,  appears elongated SE-NW at the lower resolutions. In the highest resolution 3.6 cm maps  it breaks up into 3 sources in a line at p.a.$\approx-40^{\circ}$.  Another distinct source lies west of the center. South of the center is a region $\approx 2\arcsec$ in length, with 3 distinct sources in a background of extended clumpy emission.   These 7 distinct sources were fit with single gaussians and their sizes, positions and fluxes are in Table 2.  We discuss the nature of these sources in the next section.
\footnotetext{This source has been identified with the galactic nucleus in earlier radio maps, and this has been assumed by other authors: \citet{2008ApJ...682.1020K}, for example, attribute the X-ray source they find near here to the 'nucleus'.  However kinematic and other considerations, discussed in section 4.3,  cast doubt on that identification, and it will be called `the central source' in this paper}.

In addition to the compact sources there is extended emission over a $6\times 6\arcsec$ region.  Figure 3 shows a naturally-weighted  3.6 cm image plotted so as to emphasize the lower level emission.  It is clearly not axisymmetric.  Weak emission connects the central and western radio source with what \citet{BE05} described  as  an `arm-like feature', and the southern and central source are immersed in what could be a nuclear spiral, a nuclear bar, or part of a starforming ring or disk.   Non-axisymmetric structure has been seen in NGC~4194  and interpreted as a bar \citep{JB06},  but that was at much larger (kpc) radii and does not seem related to the nuclear emission.   These structures are discussed in $\S$4. 

\subsubsection{Thermal Emission and the Spectral Index}

The radio continuum of starburst galaxies combines non-thermal emission from synchrotron sources created in previous generations of stars with thermal free-free emission generated in  young {\HII} regions tracing the current episode of star formation \citep{{Co92},{TH94}}.  Non-thermal processes typically dominate the longest wavelengths (20 cm and longer) and thermal free-free
 the shorter 2~cm and mm bands.  
The radio spectrum can be characterized by the spectral index $\alpha$, such that $S_\nu=\nu^{\alpha}$:  for pure thermal free-free emission from {\HII} regions, $\alpha = -0.1$, for optically thick emission $\alpha > -0.1$ and can be as great as $+2$.  For non-thermal synchrotron emission 
$\alpha$ is steeper, i.e. more negative, than the thermal value.   Non-thermal emission of starburst galaxies is observed to  have $\alpha$ in the range $-0.5$ and $-1.2$ between 21 and 2 cm.  The integrated 21 cm flux of \citet{CO90} and the single dish 6 cm result of \citet{BE91} give the spectral index of NGC 4194 as $-0.81$ between 21 and 6 cm. 

The archival 2~cm  observations offer a wider baseline view  of the spectral index.  We created a  2 cm image with the same beam size as the 6 cm and took the spectral index between 6 and 2 cm, which is shown in Figure 4 along with the 2 cm  map contours.    The 6 to 2 cm spectral index shows that the central and southern sources have substantial non-thermal emission and the western is purely thermal. In fact, the spectrum of the western source may actually be rising at 2 cm, consistent with a thermal source of such high emission measure as to be partly optically thick at 6 cm.  

The radio continuum at 6 and 3.6~cm  
is a mixture of thermal free-free emission
and non-thermal synchrotron emission, but the 2 cm map is mostly thermal.  This is shown by comparing the fluxes to those at 2.7 mm, where the non-thermal emission is insignificant
and dust emission is negligible.  
 \citet{AB10} measured $4.6\pm1$ mJy 
 total flux at 2.7 mm over this entire region.  With $\alpha=-0.1$ this predicts total {\it thermal} fluxes  $5.6\pm1$ mJy at 2 cm and  $6.3\pm 1$ mJy at 6 cm. 
 So 88\% of the total 2 cm flux is thermal (but only 15\% of the total 6 cm.) 
 The 2~cm emission in Figure~2 is thus an excellent high-resolution
 representation of the extinction-free
 star formation traced by the free-free emission. 
 
  The Lyman continuum rate to maintain this 
 thermal flux is estimated from the prescription of \citet{HBT90} and the 3mm flux to be  $N_{Lyc} = 1.0 \times 10^{54}~(D/\rm 39~Mpc)^2~s^{-1}$ for the three main star-forming
 regions of the central 
 $\sim$10\arcsec\ . 
 This corresponds
 to $L_{OB} = 3 \times 10^{10}~\rm L_\odot$, and a star formation rate of 
 $\sim$10~$\rm M_\odot~yr^{-1}$ for a Kroupa 3 Myr starburst based on STARBURST99.

The non-thermal emission in the compact sources of NGC 4194 is probably from recent supernovae.   There must have been considerable supernova activity in the recent past:  the 3.3 mJy of 3.6 cm emission which we calculate to  be non-thermal requires ~450 objects like CasA.  \citet{BE05} found that at 21 cm, where the non-thermal component dominates,  the emission appears with  a MERLIN beam of $0.17\times0.15$\arcsec~  to be a very compact double source.  The two components have 12.6 and 5.1 mJy and their positions and separation are consistent with Central Sources 1 and 3 in our 6 and 3.6 maps.  There is also a weak 21 cm source north of the midpoint of the double source with peak flux density 0.7mJy/bm; it has no 3.6 cm counterpart.  We cannot, with our present data, map the spectral index with sufficient spatial resolution to locate the non-thermal sources precisely.  The 6 to 2 cm spectral index hints that  the emission is more strongly non-thermal on the northern side of the central source, perhaps related to the weak third source in the 21 cm map, but that is not conclusive: the 2 cm data were observed in a different epoch than the 6 cm and the precessesion may not be perfectly accurate. 
We conclude that thermal and non-thermal emission are mixed in the central and southern sources of NGC 4194 at wavelengths 3.6 cm and longer.  

The purely thermal index of the small western source argues that it is a young  \HII~region that has not yet produced a  significant number of supernovae.   From the radio flux we deduce that the Lyman continuum flux in that source is $N_{Lyc}\approx 2.08\times10^{53}~ \gamma~s^{-1}$,
equivalent to $2.08 \times 10^{4}$  O7V stars and a star formation
rate of $\sim 2~\rm M_\odot~yr^{-1}$.   
If this is one star cluster it is one of the largest super star clusters yet seen; it is much brighter than the central source in NGC 5253 and almost as strong as the SW radio source in NGC 4102 \citep{BT10} and the brightest source in M82 \citep{2009AJ....137.4655T}. 
The extended nature of the western source argues that it is probably more than one cluster, however. 

In summary, the three main radio sources are intense regions of star formation, but at different evolutionary stages.  The central and southern source are old enough that many of their stars have gone supernova.  The presence of an X-ray source (presumably an X-ray binary) near the central radio source \citep{2008ApJ...682.1020K} is another sign of stars in very late stages of evolution.  The western source, in contrast, is too young to have formed a significant number of supernovae. 

\subsection{[NeII] Spatial Structure Agrees with the Radio}

The spatial distribution of the [NeII] emission is found from the data cube by collapsing the cube along the velocity axis.  The resulting map is shown in Figure 5, superimposed on a radio map convolved to the same beam size.  In the [NeII]  map the western source is a discrete peak, while the southern source is a plateau extending the central source; the radio continuum at the same spatial resolution looks very similar. [NeII] is a star formation tracer showing the presence of young {\HII} regions, so its distribution here argues that all the nuclear sources, including those with non-thermal radio spectra, host some recent star formation.

The total flux of [NeII] summed over the line extent and the region mapped is 
$1.6\times10^{-12}\rm ~erg~s^{-1}~cm^{-2}$, with absolute uncertainty $\approx30\%. $ 
The IRS on Spitzer   had two pointings  of the SH module on this position and measured $4.7$ and $5.0$ Jy, which is  $2\pm0.06\times10^{-12}\rm ~erg~s^{-1}~cm^{-2}$.  There are considerable systemic uncertainties in converting Jy to flux for an unresolved line and in comparing observations of such different resolution as TEXES and IRS; we can say only that the results are consistent with TEXES having recovered the entire Spitzer flux.  

\subsubsection{$[Ne^+]$ Abundance} 

The strength of the [NeII] line and of the thermal radio continuum are related because both depend on the emission measure $n_e^2\ell ~\rm (cm^{-6}~pc)$  If the neon abundance has the solar value of  $8.3\times10^{-5}~\rm n(H)$ by number, and is all in  $Ne^+$ (an overestimate, since Spitzer detected [NeIII]), the collision strengths of \citet{OF06} give
$F([NeII])(erg s^{-1}cm^{-2})=2.0\times10^{-10}F_{5 GHz}\rm (Jy)$.   This predicts a total 5 GHz flux of 8~mJy for NGC 4194, compared to the 6~mJy deduced above for {\it thermal} radio emission.  The sizable uncertainty in our absolute flux calibration permits us to say only that the flux is consistent with a solar or somewhat higher abundance, and even marginally consistent with the slightly less than solar metallicity cited by \citet{OH06}. 

 \subsection{Ionized Gas does not Coincide with the  Molecular Gas }
  \citet{AH00} used the OVRO interferometer to map the $^{12}CO(1-0)$ emission in the center of NGC 4194 with beam sizes 1.7\arcsec--2.5\arcsec\ and velocity resolution of 20\kms.  They found that in the region mapped in  radio and [NeII]  the molecular gas comprises 5 distinct peaks and  an extended component.   We precessed their  figures to the J2000 system used in the radio continuum observations and show the 2~cm map overlaid on the CO in Figure 6.  Since both the 
radio and CO maps are interferometer maps, the uncertainty in registration should be
negligible. 
It is clear from the figure that {\it the spatial distribution of molecular gas differs materially from the ionized}.  First,  the brightest radio and [NeII] source is not perfectly coincident with any of the 5 CO peaks, but sits on the north-west side of CO peak B.  The offset is $\approx0.6\arcsec$,  
much larger than the 0.1\arcsec~ conservatively estimated for the possible error in the precession.   
  Second,  the southern clump of radio sources agrees in position with the molecular peak A,  but not in relative intensity: the radio and [NeII] flux on  peak A is much weaker than on peak B,
  which has the brightest radio emission.  
  The CO/2 cm ratio on A is higher than on B by at least a factor of 5.   Third,  there is no ionized gas detected on peak D even though it is almost (80\%) as strong in CO as peak C, which is coincident with the western radio source; the CO/2 cm ratio in D must be at least 10 times higher than in C.     
  
  We see that the greatest concentration of molecular mass is neither the radio peak nor the geometrical center of the starburst activity, and that star formation activity is concentrated on one side of the molecular gas distribution.   In other words, the ionized gas does not trace the mass distribution on scales of $\sim$0.5\arcsec\ or $\sim$~80--100~pc.  The mass of young stars in the  \HII~regions is not significant on these scales:  the total stellar mass in the clusters is less than a few times $10^7M_\odot$, a small fraction of the total molecular mass of $\approx2\times10^9 M_\odot$ and of the $\approx10^8 M_\odot$ in each cloud.   
  
This raises a new question: {\it where and what is the galactic nucleus?}.   We cannot assume that the nucleus is identical with one of the radio peaks--not even the central source.   We return to this in the Conclusions. 

 \section{Kinematics}
 
 The kinematics of molecular gas in NGC 4194 were studied by \citet{AH00}, who observed CO emission with a $2.0\times 1.7$\arcsec~beam and 20\kms~ spectral resolution, and by \citet{AB10} who measured $^{13}$CO with a $4.56\arcsec\times3.98\arcsec$ beam. The atomic gas was observed  by \citet{BE05}, who had 21\kms~resolution.  The velocity field in NGC 4194 is extremely complex, reflecting the merger that created the galaxy. \citet{AH00} and \citet{BE05}  agree that in the inner  $\approx0.65 kpc$ (3.5\arcsec) there is a shallow  N-S velocity gradient of ~$\approx 60 $~\kms per arcsecond,  consistent with rotation, that there are significant non-circular motions at larger distances, and that the nuclear region contains some velocity components not due to rotation.   
 
\subsection{Kinematics of Ionized Gas: [NeII] Distribution}
The TEXES observations are the first to measure ionized gas kinematics in NGC~4194.   The spatial distribution of the emission is shown as a 2-D grid of spectra in Figure 7a. The [NeII] emission in the central $5\times5$\arcsec~ covers the velocity range $2375-2650$\kms.  This is close to the full range over which \citet{BE05} see HI absorption in the central  $3$\arcsec.  The overall distribution of the peak velocities also agrees roughly with the HI result:  the highest velocities observed are around $2600$\kms~ and  are in the southern extension, the lowest  are around $2400$\kms~ and  are at the north end of the source , and the north west is at intermediate velocities of $\sim2500$\kms.  

Besides the similarities there are substantial differences between the ionized gas kinematics, as shown on the spectral grid, and the atomic and molecular findings.   While $Ne^+$ is the commonest ionization state of neon in a high-metal star formation region like NGC 4194 and is probably widely distributed in the galaxy, the [NeII] line is collisionally excited and is most readily seen in \HII~regions of moderate to high density (i.e., $\gtrsim 10^3 ~\rm cm^{-3}$).  So the [NeII] is concentrated in the compact and intense radio sources to which the high resolution radio maps are sensitive.   The spatially isolated $2500$\kms~ feature may be straightforwardly associated with the western thermal source.  We assign the $2440$\kms~ feature to the nuclear source and the $2600$\kms~ to the southern emission complex.  The spatial resolution is not high enough to separate the central and southern source completely and they overlap in some positions, creating very complex line profiles (as for example, in Figure 7b). 

\subsection{[NeII] Line Profiles}
What can the [NeII] spectra tell us about the ionized gas motion in this galaxy?  The profile of a fine-structure emission line can be formed and influenced by thermal velocity dispersion, gravitational turbulence in a virialized system (such as a bound nebula), and non-turbulent bulk motions.  The obvious bulk motion in NGC~4194 is rotation creating a velocity gradient. Does this contribute to the line profiles, or can we treat the sources as  kinematically isolated?  We explored this question by creating data cubes from the spatial structure of the radio maps and a given velocity gradient.  For a reasonable gradient like the $\approx 60 $\kms~arcsec$^{-1}$~of  the CO, the gradient does not affect the line widths of the individual sources.  The sources are simply too small to include a significant range of the gradient velocities.  This justifies analysing the line profiles in terms of internal, rather than large-scale galactic, motions. 

Thermal velocity dispersion is not an important factor in these line profiles; because of the high mass of neon
 an  \HII~region with $T_e =7500 K$ has FWHM from thermal effects of only $4.1$\kms~. The FWHM increases only as the square root of $T_e$.  Gravitational turbulence in a virialized system will create a gaussian line whose FWHM can be found from $\sigma=FWHM/2.35= (GM/3R)^{1/2}$ where $R$ is the radius of the system.  But since we do not resolve the individual emitting regions in the [NeII] sources we do not know $R$. Further, the radio maps show that the central and southern [NeII] source each include several distinct radio sources, presumably super star clusters, whose sum creates the observed lines.  So while the gravitational turbulence is probably responsible for most of the width of the line profile, it cannot be satisfactorily calculated from these data for the southern and central sources.  For the spatially isolated and apparently simple western source, we show the line summed over a $3\times3$ pixel box in Figure 8 along with the best single gaussian fit.   The FWHM is 104\kms~ and line center is $2480$\kms~;  $\chi^2/d.o.f.$ is 1.03 and the residuals do not show structure.  The virial velocity relation formally gives $M_\odot/R(pc) \approx 1.4\times10^6 M_\odot/pc$, while the total stellar mass estimated from the $N_{lyc}$ and a Kroupa IMF extending to $0.1M_\odot$ is only $8\pm3\times10^5 M_\odot$. The source is unresolved with the  0.18\arcsec  ($\approx 35~pc$) beam of the highest resolution radio map.   It is likely that even this apparently simple source contains multiple clusters at offset velocities, which make the line appear wider than the true turbulent FWHM. 
Line profiles in the central and southern sources have FWHM from $85--104$\kms~, which formally gives $M/R$ of $0.8-1.4\times10^6M_\odot/pc$, but since these sources clearly (from the radio maps) contain multiple sub-sources both the FWHM and the mass concentrations are only upper limits.   

\subsection{ [NeII] Kinematics on Large Scales: Position-Velocity Diagrams}
To examine what  [NeII] can tell us about the large scale kinematics of the nuclear region, we created a data cube binned by 4 spectral pixels and 2 pixels along the slit, to create almost square $0.7\arcsec \times0.7\arcsec\times3.74$\kms~pixels.  From this we can see the ionized gas kinematics with minimal spatial averaging.   Figure 9a shows the line extent as a function of position along a N-S cut through the central source;   Figure 9b shows an E-W cut,  also through the center. 
 The N-S position-velocity diagram has an obvious velocity gradient  of several tens of \kms~arcsec$^{-1}$~with blue north and red south, very similar to the CO \citep{AH00} and HI \citep{BE05} results.   The ionized gas gradient is  $200$\kms~in 4.2\arcsec, for a spatial gradient (corrected for the $40^{\circ}$ inclination angle)  of 390 \kms kpc$^{-1}$. The total velocity extent is similar to the CO results but the gradient is slightly higher than the 320 \kms kpc$^{-1}$ of the atomic and molecular gas; this may be because of the very different beam sizes.  
 
The [NeII] observations average over a very patchy distribution of ionized gas. We checked  that this does not distort the position-velocity diagram by comparing it to that of the simulated data cube based on the radio continuum maps;  they are consistent.  {\it The gradient is not due to averaging over sources at offset velocities: the velocity offsets between the [NeII] sources result from their positions along an incompletely sampled velocity gradient.}

  In the E-W position-velocity diagram the central and the western source appear spatially separated and with almost identical velocity ranges of 2400--2550\kms~. There is nothing to suggest an E-W gradient;  we find from the simulated cube that  if there were a gradient as large as 20 \kms~per arcsecond, it would appear in the position-velocity diagram.   There is a separate red feature at 2580\kms~, most clearly in the central source; we believe that to be part of the southern source which was included in the same beam. 
   
  \subsection{A Galaxy Core in Solid-Body Rotation}
   In the previous sections we showed that the none of the  \HII~regions measured in the [NeII]  and radio are likely to be a galactic nucleus. But even if the mass distribution is not apparent in the CO, radio and [NeII] maps, its gravitational effects cannot be hidden.  We now find that there is no {\it kinematic} sign of a nuclear mass concentration either.   In a spiral galaxy there is usually a  very broad line on the nucleus and a velocity jump across it;  there is no such pattern  in NGC 4194.  Specifically the central source, the apparent candidate for nucleus, has the  {\it narrowest} line profile of all the [NeII] sources.  The only kinematic hint at a mass concentration is in the CO velocity field; the isovelocity contours, which are evenly spaced across the main radio and CO peaks, kink and take on a closer spacing around $\alpha=12^h14^m09.5^s, \delta= +54^o 31^{'} 36^{''}$. But that location is between the central and western radio sources and there is no emission peak, at any wavelength, there.   Instead, the kinematics are dominated by  a smooth gradient N-S in both molecular and ionized gas.   Smooth gradients imply solid body rotation.  In disk galaxies solid body rotation is usually associated with rings, and elliptical galaxies, especially those that are post-merger, may host cores several hundred pc in size, kinematically distinct from the rest of the galaxy, with this rotation pattern \citep{TF91}.  
   
How can the galaxy have the kinematic signature of solid body rotation while the ionized gas appears to be a two-armed spiral?     If we critically examine the low-level radio emission maps in Figure 3,  it is clear that the appearance of `spiral structure'  depends very strongly on the southeast  elongation of the southern source.   The sub-sources fall on a line southeast-northwest and the emission extends further in a spur in the same direction. The  whole southern complex is almost perpendicular to the emission connecting the southern and central sources, so that together with the elongated central source the region suggests a  `twin peaks'  star formation structure--but again, that contradicts the kinematics, which are not at all like those of the barred spirals where `twin peaks' form.  Finally, the  `arm-like' arc of radio emission connecting the central and western sources lies north-west of the main CO emission in the region of highest radio-to-CO ratio;  it may show a front of star formation proceeding into the clouds, rather than a true spiral arm.  
So the spiral structure which appears to first glance is less certain on closer inspection. 

Let us now work from the opposite direction: how does the ionized gas distribution fit the kinematic results?  The central and southern sources are clusters that have formed in a rotating spheroidal core with southern end receding.  We cannot rule out that the rotating region is a very thick ring or disk instead of a spheroid; if this case the clusters have formed throughout the thickness of the ring.  That the western source has almost the same velocity as the central may agree with the finding of  \cite{AH00} that the velocities of cloud C and D are consistently lower than if they were in solid body rotation with A and B.  The western source, like its host cloud, may not have yet settled into equilibrium.

     \section{Conclusions}
We have presented high-resolution radio continuum maps at 6, 3.6 and 2 cm, and a high spectral-resolution velocity-position cube in the [NeII] $12.8\mu$m emission line, of the central $1.3~kpc$ of NGC 4194.  These maps are the first high resolution, extinction-free, measurements of ionized gas in the center of this galaxy and trace the star formation regions with resolution up to 0.18\arcsec  ($\sim35~pc$).  We find that: 
\begin{itemize}
\item NGC 4194 hosts a nuclear starburst that has formed multiple compact sources, apparently groups of embedded super star clusters,  within a ~3\arcsec ~radius.   The low-level radio emission suggets a 'nuclear spiral' but that is not confirmed by the kinematics. 

\item Based on the thermal radio emission at 2 cm, which agrees well with previous observations
at 2.7mm, we find  for the central $\sim$10\arcsec\ region a thermal free-free flux of 
 $S_{2cm}^{thermal} = 5.6 \pm 1~\rm mJy$, which implies a Lyman continuum
rate of $N_{Lyc} = 1.0 \times 10^{54}~(D/\rm 39~Mpc)^2~s^{-1}$. 
For a Kroupa IMF and a 3~Myr starburst,
this implies a star formation rate of $\sim$10~$M_\odot~\rm y^{-1}$, and a luminosity in massive
young stars of $L_{OB} = 3 \times 10^{10}~\rm L_\odot$, about a third of the total infrared
luminosity of $L_{IR} = 8.5 \times 10^{10}~\rm L_\odot$. The star formation in
this galaxy is dispersed widely over the inner 10\arcsec~(1.9 kpc) region. Our radio 
observations are not sensitive to regions with 2 cm fluxes less than $\sim 0.1$~mJy,
or $N_{Lyc} < 2 \times 10^{52}~(D/\rm 39~Mpc)^2~s^{-1}$, so it is possible that the 
remaining infrared luminosity arises in smaller, and widely dispersed, star forming regions within
the system.

\item  Of the compact emission sources that are bright enough to analyze,  one has a pure thermal or slightly rising spectrum, typical of a very young embedded \HII~region. The others are older and have a mix of thermal and significant non-thermal  emission.     The [NeII] line velocity dispersions are consistent with the gravitational effects of  many dense star clusters.   
\item The ionized gas has a smooth N-S velocity gradient of 390 \kms kpc$^{-1}$  across the observed region, consistent with a core in solid body rotation.  
\item  None of the radio sources can be convincingly identified with a galactic nucleus. The nuclei of the original  merger partners have not been identified.
\item The CO velocity field shows that a mass concentration is present at $\alpha=12^h14^m09.5^s, \delta= +54^o 31^{'} 36^{''}$.  This location is in between and roughly equidistant from the three star formation regions, and there is no star formation detected there.   

\end{itemize}

 How can the last two items on this list be reconciled with the scenario that the Medusa is a minor merger?  Each of the merger partners presumably hosted a nucleus of mass $10^6-10^8M_\odot$ that would be detected via their influence on the kinematics.  It is possible, and would be consistent with the simulations, to have both original nuclei now included in the mass concentration detected in the isovelocity contours.  But in this case the nuclei must be devoid of star formation and ionised gas, or they would be detected in the radio and [NeII].   The gas and stars in the center of NGC 4194 have not come to equilibrium and the starburst may continue in regions now quiescent.   Measurements of ionised and molecular gas with higher spatial resolution, and simulations exploring the central kpc,  could perhaps find more clues to the behaviour of this remarkable galaxy. 
 
 \bigskip
 \bigskip
TEXES observations at the IRTF were supported by NSF AST-0607312 and by AST-0708074 to Matt Richter. This research has made use of the NASA\&IPAC Extragalactic Database (NED) which is operated by the Jet Propulsion Laboratory, Caltech, under contract with NASA.  We thank an anonymous referee for careful and thoughtful comments.

\clearpage

\begin{table}
\caption{Observations and Reductions}
\begin{tabular}{ccrrrrrrrrcrl}
\tableline\tableline

Date & Instrument & Wavelength & Beam Size &$\Theta_{las}$\tablenotemark{a} & noise & Program \\
\tableline
2/2/2013 & TEXES & $12.8$\um & n.a. & n.a.& 1.4\arcsec & n.a.\\
6/7/1987 & VLA & 2 cm &$ 0.18 \times 0.149 $ \arcsec & 3.6\arcsec& 0.06 mJy/bm & AE47\\
21/11/1987 & VLA & 6 cm &  $1.4\times 1.13$\arcsec & 24\arcsec&  0.05 mJy/bm &  AS286\\
20/11/2000 & VLA & 6 cm & $0.27 \times 0.17$\arcsec & 8.9\arcsec &0.08 mJy/bm & AN095\\
 " & " & " &$0.46 \times 0.43$\arcsec & 8.9\arcsec &  0.02 mJy/bm & "\\
" & " &3.6 cm  &  $0.18 \times 0.104$\arcsec & 5.3\arcsec & 0.023 mJy/bm & "\\
" & " & " & $0.26 \times 0.23$ \arcsec & 5.3\arcsec &  0.015 mJy/bm  & " \\
19/3/2001 &"  & " & $0.53 \times 0.53$\arcsec & 17\arcsec & 0.04mJy/bm & " \\
\tableline
\end{tabular}
%% Any table notes must follow the \end{tabular} command.
\tablenotetext{a}{$\Theta_{las}$ is the angular size of the largest structure to which the observations were sensitive.}
\end{table}
%\end{}

\begin{table}
%\begin{center}
\caption{Source Parameters from 3.6 cm Maps\tablenotemark{a}}
\begin{tabular}{crrrrrrrrrrr}
\tableline\tableline
ID& RA (2000)& Dec (2000) & Major Axis & Minor Axis & Peak Intensity & Total\\ 
\tableline
Central 1 & 12 14 09.68 & 54 31 35.73 & 0.46\arcsec & 0.22\arcsec  & 0.8 mJy/bm &1.2 mJy\\
Central 2 & 12 14 09.67 &54 31 35.74  & 0.21\arcsec & 0.14\arcsec& 1.16 mJy/bm & 2 mJy \\
Central 3 & 12 14 09.65 & 54 31 35.86 & 0.22\arcsec & 0.11\arcsec  & 1.9 mJy/bm & 3.3 mJy \\
\tableline
W. Source & 12 14 09.326 & 54 31 35.68 & 0.23\arcsec & 0.14\arcsec & 0.46 mJy/bm & 0.94 mJy\\
\tableline
South 1& 12 14 09.59 & 54 31 34.6 & 0.5\arcsec & 0.37\arcsec & 0.3 mJy/bm & 0.57 mJy\\
South 2 & 12 14 09.63 & 54 31 34.27 & 0.51\arcsec & 0.43\arcsec & 0.46 mJy/bm & 0.9 mJy\\
South 3 & 12 14 09.66 & 54 31 34.21 & 0.69\arcsec & 0.513\arcsec & 0.46 mJy/bm & 1.24 mJy\\
\tableline
\tableline
\end{tabular}
%% Any table notes must follow the \end{tabular} command.
\tablenotetext{a}{The nuclear and western cluster fits are from the uniform weight map with a $0.18\times 0.1$\arcsec ~beam; the southern clusters are fit from the natural weighted map with a $0.26\times 0.23$\arcsec ~beam}
%\tablenotetext{b}{Yet another sample footnote for table~\ref{tbl-2}}
%\tablenotetext{c}{Another sample footnote for table~\ref{tbl-2}}
%\tablecomments{We can also attach a long-ish paragraph of explanatory
%material to a table.}
%\end{center}
\end{table}
\clearpage

\begin{figure}
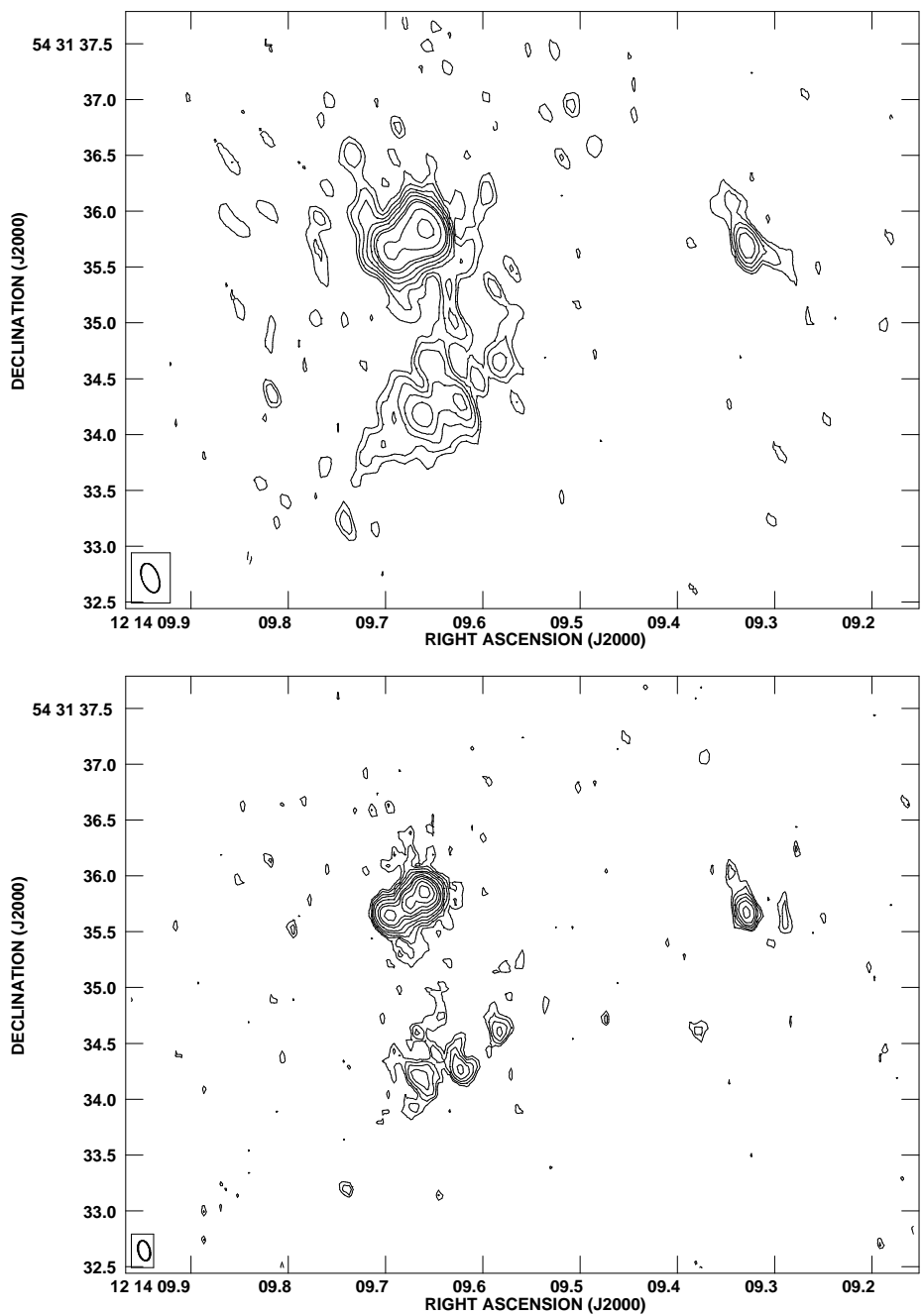

\begin{center}
\includegraphics*[width=3.5in, angle=-90]{FIG1A.EPS}
\includegraphics*[width=3.5in, angle=-90]{FIG1B.EPS}
\caption{The uniform-weight maps at 6 (top) and 3.6 cm (bottom).  For the 6 cm map, the B and A array observations were combined to create a data set with a  $0.28\arcsec\times 0.19\arcsec$ beam, shown in lower left,  which was sensitive to structures as large 24\arcsec.  The lowest positive contour is $2.5\sigma = 0.095~{\rm~ mJy}$.  The 3.6 cm map has a $0.18\arcsec\times 0.1\arcsec$ beam, shown, and the lowest positive contour is $2.5\sigma = {\rm 0.06~mJy}$. For both maps the levels are spaced logarithmically by factors of $2^{n/2}$.}
\end{center}
\end{figure}

\begin{figure}
\begin{center}
\includegraphics*[width=4in, angle =-90]{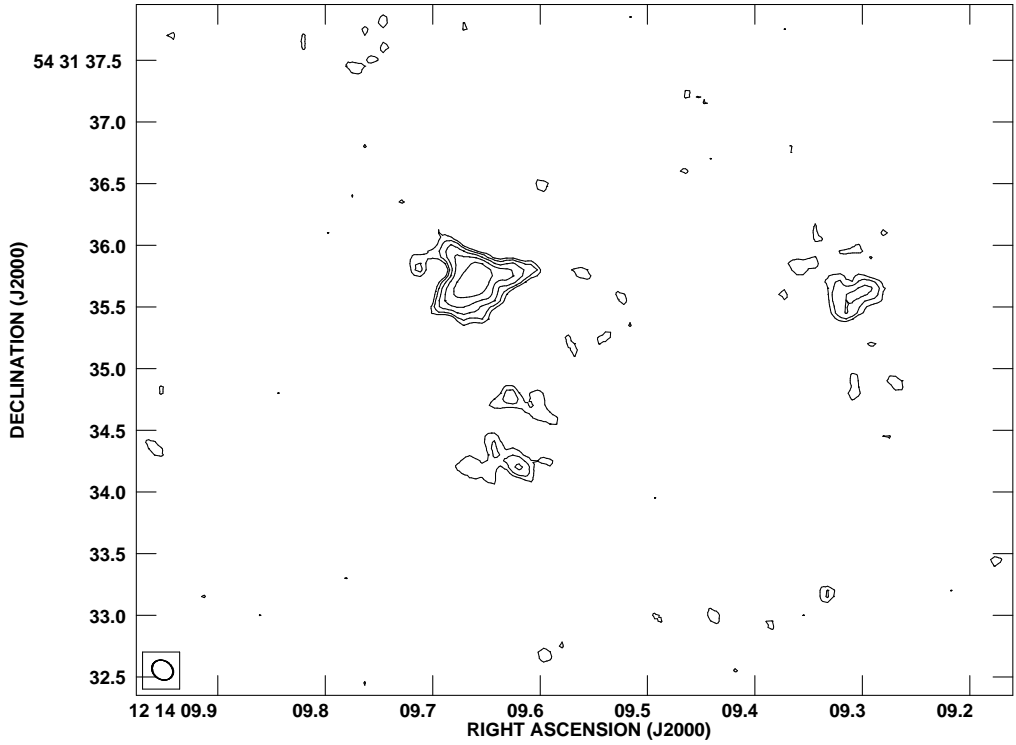}
\caption{The uniform-weight 2 cm map. The $0.18\arcsec\times0.15\arcsec$ beam is shown at lower left, The lowest positive contour is $0.15~{\rm mJy} = 2.5\sigma$ and the levels are spaced logarithmically by factors of $2^{n/2}$. }
\end{center}
\end{figure}

\begin{figure}
\begin{center}
\includegraphics*[width=3.5in, angle=-90]{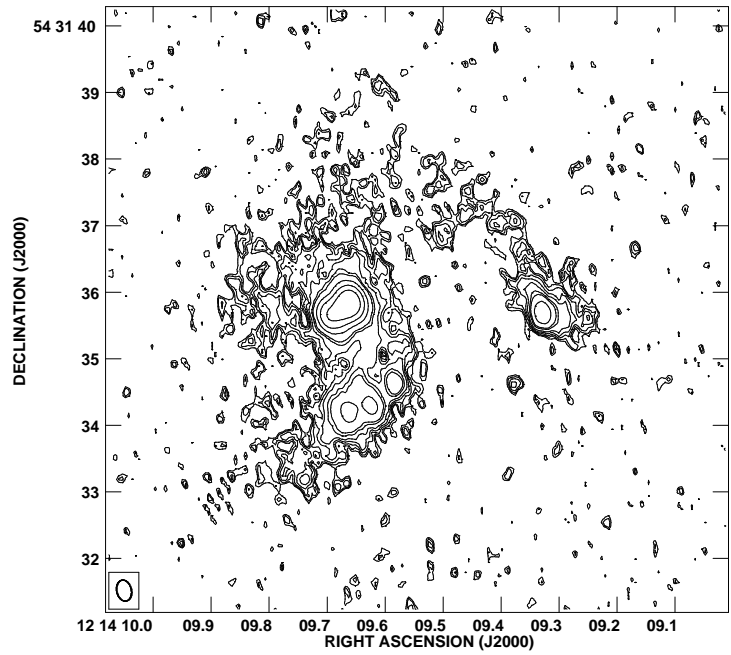}
\caption {The low-level emission at 3.6 cm. The $0.32\arcsec\times 0.22\arcsec$ beam is shown.  The lowest contour is $1.3\sigma$ (0.02 mJy/bm) and contour levels are 1 1.4 2 2.8 4 6.25 8 16 32 64. }
\end{center}
\end{figure}

\begin{figure}
\begin{center}
\includegraphics*[width=4in, angle=-90]{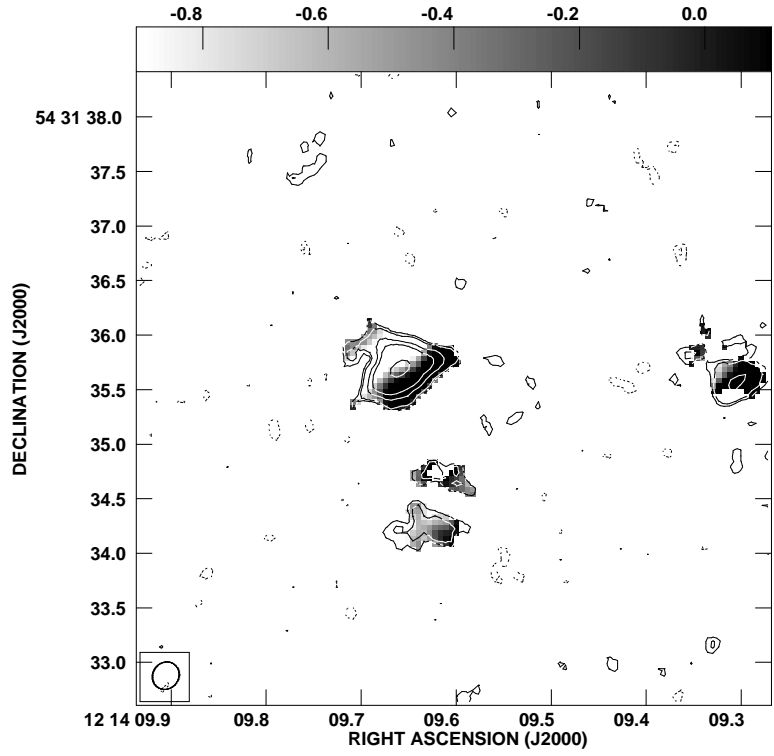}
\caption{The spectral index between 6 and 2 cm in greyscale with the 2 cm intensity in contours. The beam is $0.26\arcsec\times0.23\arcsec$ and contour interval and levels as in Figure 3. }
\end{center}
\end{figure}

\begin{figure}
\begin{center}
\includegraphics*[width=0.7\columnwidth]{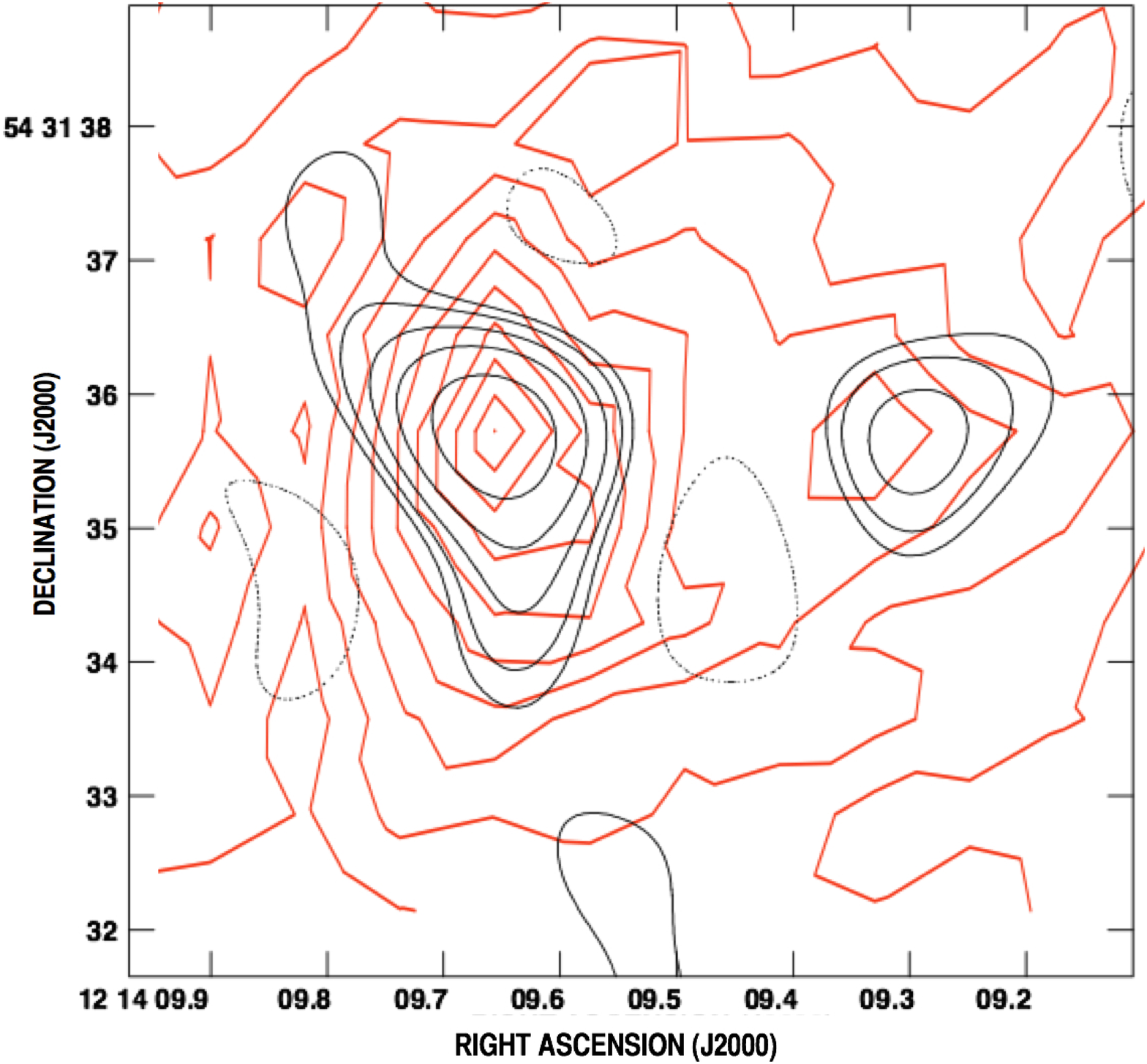}
\caption{Spatial distribution of the total line flux in the [NeII] data cube, in red.  Contours are integer multiples of $10\%$ of the peak flux of ${9\times10^{-3}~\rm erg(s~cm^{2}~sr)^{-1}}$. The apparent peak at RA 12 14 09.9 is spurious,  from a flaw in the detector.  Black contours show the 2 cm radio emission convolved to the [NeII] beam; contours are intergral multiples of 1.68 mJy/bm and dashed contours are negative. }
\end{center}
\end{figure}

\begin{figure}
\begin{center}
\includegraphics*[width=1\columnwidth]{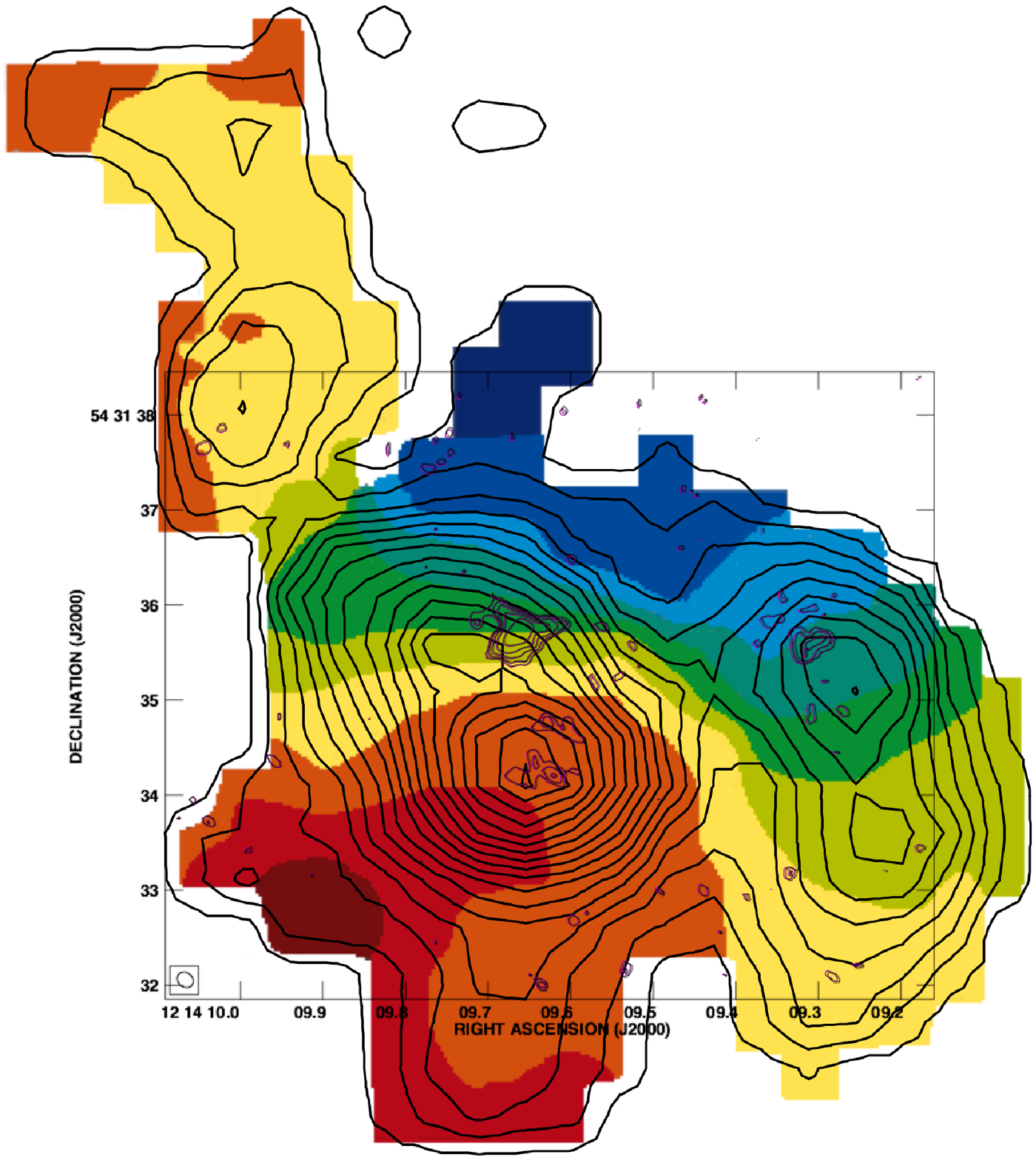}
\caption{The radio continuum contours of the VLA 2~cm map are superimposed on the CO integrated intensity and CO velocity field from \citet{AH00}.  The estimated accuracy of the overlay is $\approx 0.2\arcsec$. The velocity contours are 2400 (blue) to 2640 (red) \kms~with 24 \kms~ spacing. The CO contours are the total integrated intensity of CO emission with levels  (0., 1.5, 3.5,...) $Jy/bm$\kms.  \citet{AH00}'s peaks A,B,C and D go clockwise from the south-east and  E is the north-east feature.}
\end{center}
\end{figure}

\begin{figure}[!h]
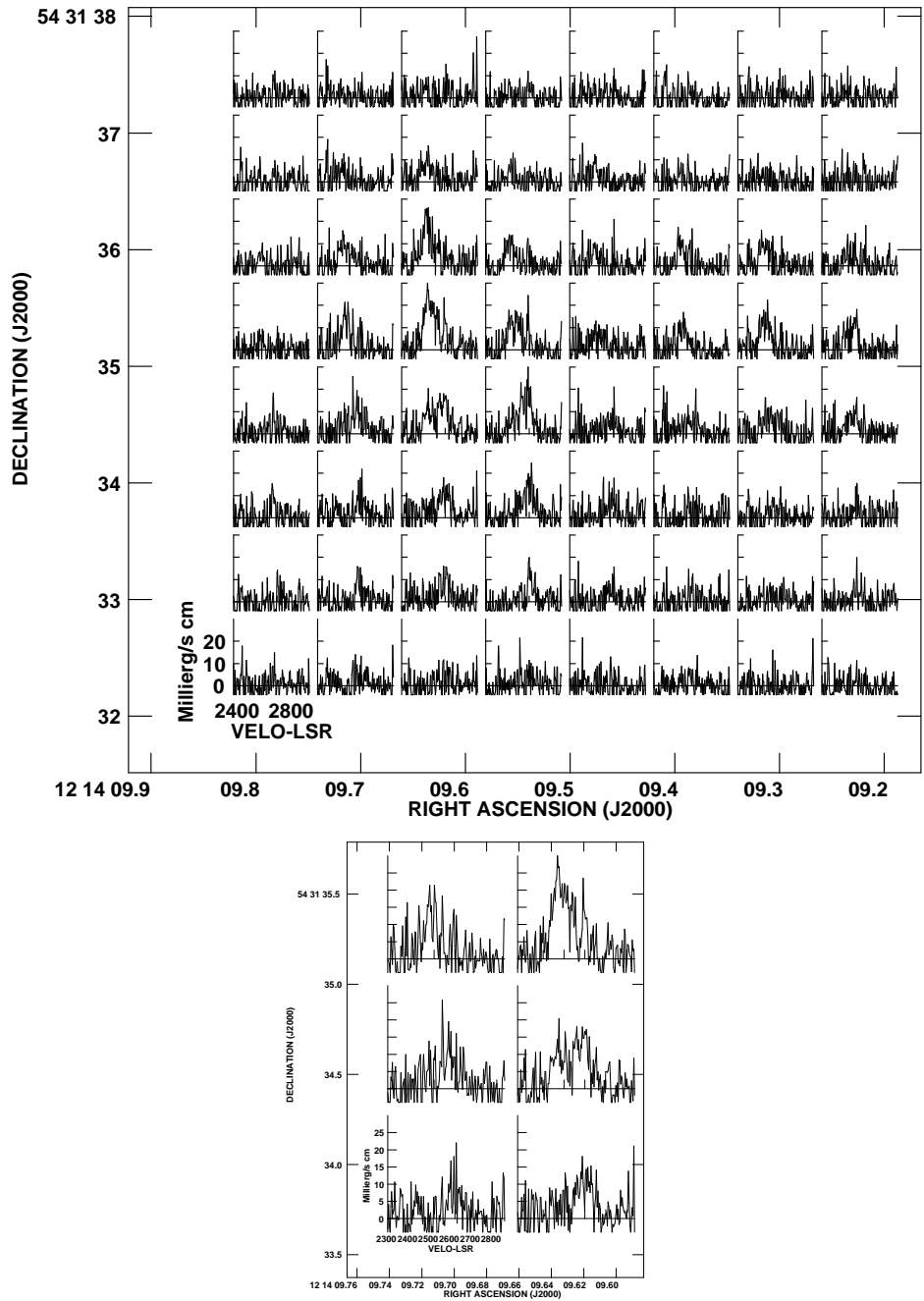

\begin{center}
\includegraphics[width=4.5in, angle=-90,clip]{FIG7A.EPS}

\includegraphics[width=2in, clip]{FIG7B.EPS}
\caption{Top: the observed [NeII] spectra in each pixel. The intensity units are $\rm erg~(s~cm^2 sr)^{-1}$. Bottom: sample pixels from the above figure which are  enlarged to show the overlapping velocity components. }
\end{center}
\end{figure}
\clearpage

\begin{figure}
\begin{center}
\includegraphics*[width=4in]{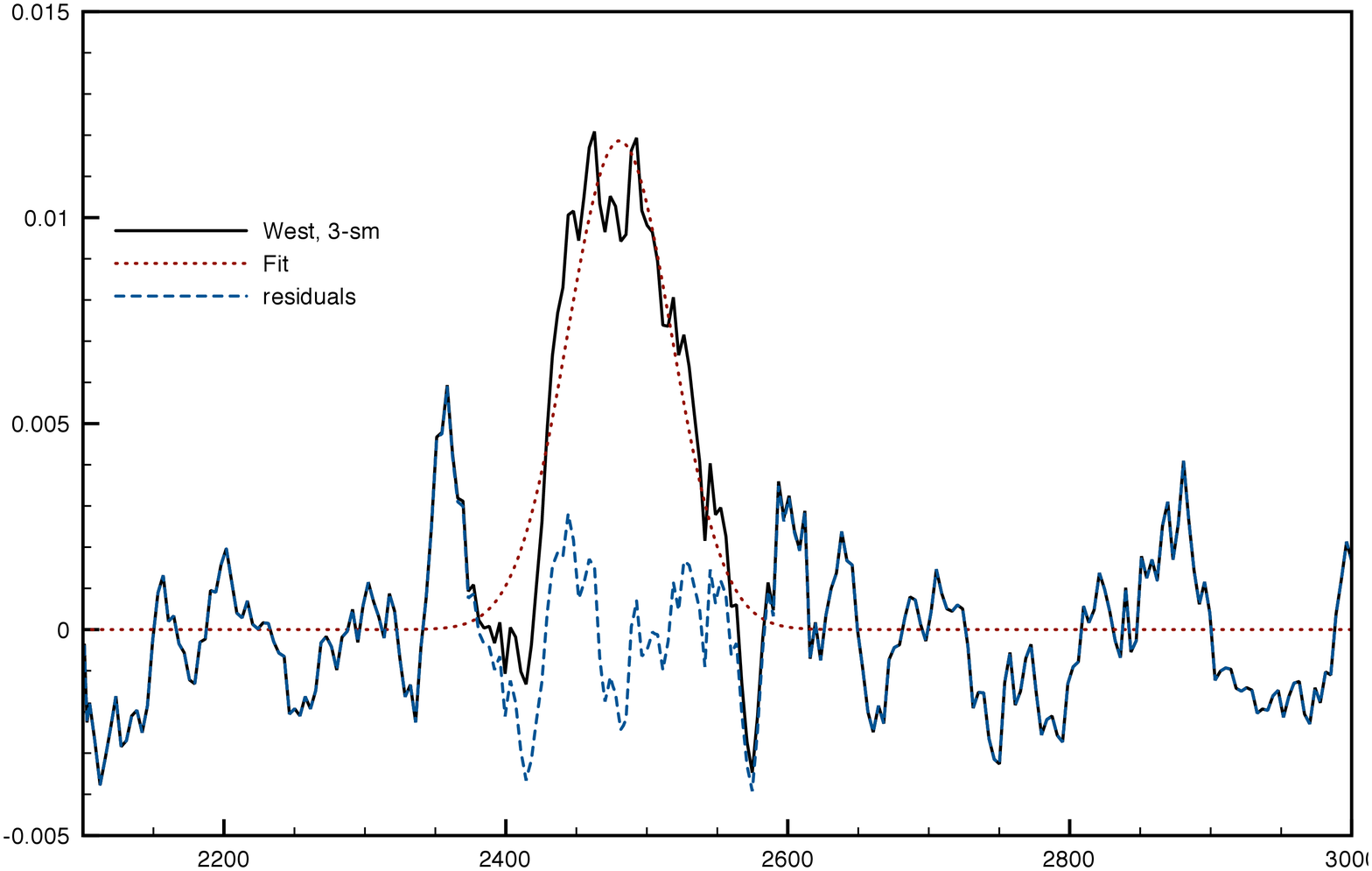}
\caption{The line profile of the entire western source, with the best single gaussian fit and the residuals.  The fit center is 2480~\kms, FWHM 103.9, and $\chi^2/d.o.f. =1.03$.}
\end{center}
\end{figure}

\begin{figure}
\begin{center}
\includegraphics*[width=5in]{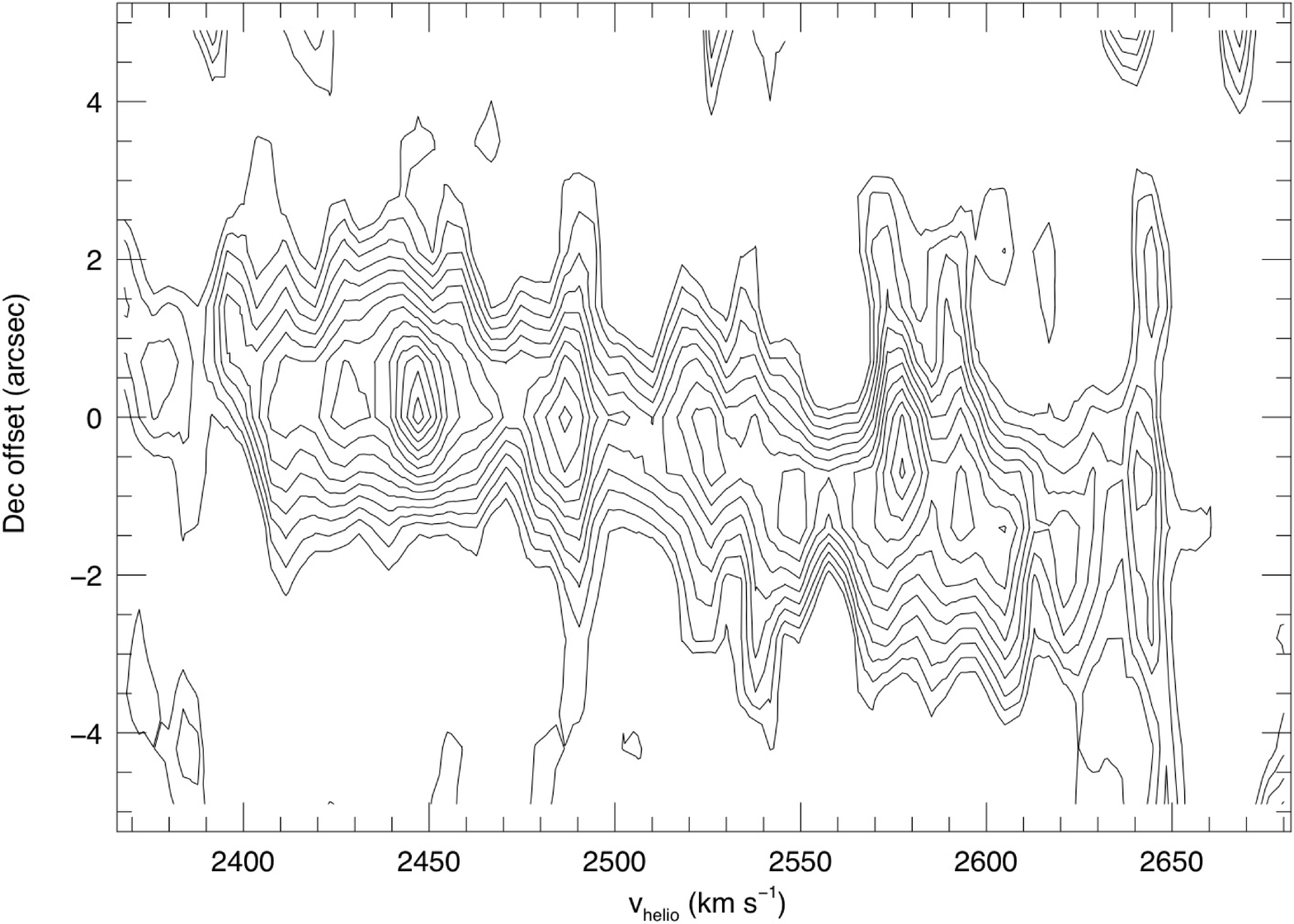}
\includegraphics*[width=5in]{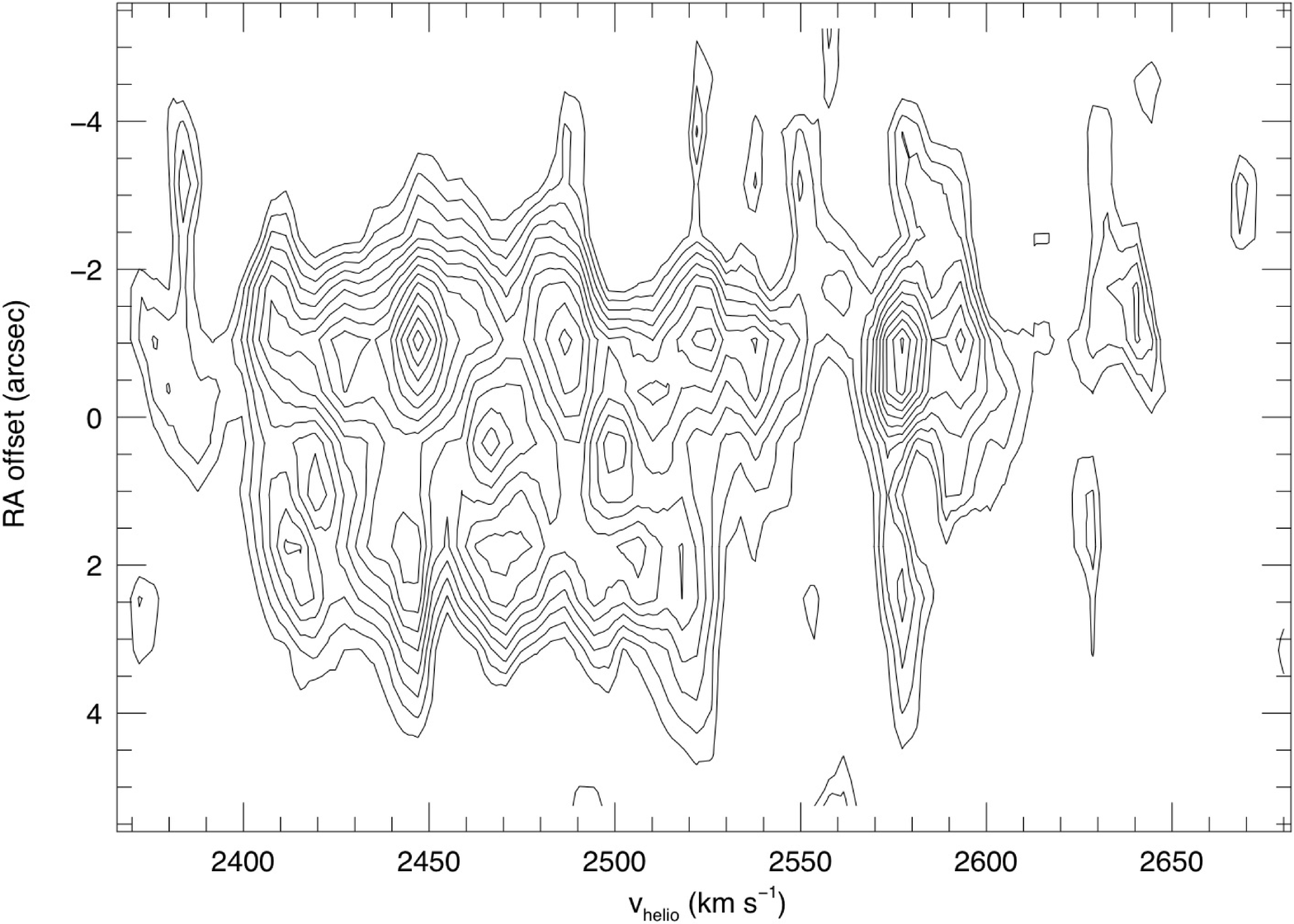}
\caption{Position-velocity Diagrams through the [NeII] sources.  For these diagrams the data was binned by 4 pixels in velocity and by 2 pixels along the slit; the final scale is $0.7\arcsec\times0.7\arcsec \times3.7$~\kms. Top is an N-S cut through the nucleus and S is up; bottom an EW cut through the brightest source with E up; in both,  blue is to the right.}
\end{center}
\end{figure}

\end{document}